# $^{19}$F nuclear spin relaxation and spin diffusion effects in the single-ion magnet LiYF$_4$:Ho$^{3+}$

B. Z. Malkin,[1a] M. V. Vanyunin,[1] M. J. Graf,[2,3] J. Lago,[3] F. Borsa,[3] A. Lascialfari,[3,4]

A. M. Tkachuk,[5] and B. Barbara[6]

[1] Kazan State University, Kazan 420008, Russian Federation

[2] Department of Physics, Boston College, Chestnut Hill, MA 02467, USA

[3] Department of Physics "A. Volta", CNR-INFM Unit and CNISM Unit, Pavia I27100, and S3-CNR-INFM, Modena, Italy

[4] Institute of General Physiology and Biological Chemistry, University of Milano, I20134 Milano, Italy

[5] St. Petersburg State University of Information Technology, Mechanics and Optics, 199034, St. Petersburg, Russia

[6] Institut Néel, Département Nanosciences, CNRS, 38042 Grenoble Cedex-09, France

**Abstract**. Temperature and magnetic field dependences of the $^{19}$F nuclear spin-lattice relaxation in a single crystal of LiYF$_4$ doped with holmium are described by an approach based on a detailed consideration of the magnetic dipole-dipole interactions between nuclei and impurity paramagnetic ions and nuclear spin diffusion processes. The observed non-exponential long time recovery of the nuclear magnetization after saturation at intermediate temperatures is in agreement with predictions of the spin-diffusion theory in a case of the diffusion limited relaxation. At avoided level crossings in the spectrum of electron-nuclear states of Ho$^{3+}$ ions, rates of nuclear spin-lattice relaxation increase due to quasi-resonant energy exchange between nuclei and paramagnetic ions in contrast to the predominant role played by electronic cross-relaxation processes in the low-frequency ac-susceptibility.

---

[a] e-mail: boris.malkin@ksu.ru

# 1 Introduction

Studies of the highly diluted paramagnet LiYF$_4$:Ho$^{3+}$ in external dc- and ac-magnetic fields are of great interest because the system can be considered a model to understand important features of quantum dynamics of the macroscopic magnetization at (avoided) energy level crossings in sweeping magnetic fields [1]. Nuclear Magnetic Resonance (NMR) has been proven to be an effective probe of the electronic spin dynamics in different paramagnetic crystals [2,3] and single molecule magnets (SMM) [4-9], and it can thus be used also in LiYF$_4$:Ho$^{3+}$. Previous measurements have demonstrated that the spin-lattice relaxation rate of the $^{19}$F nuclear magnetization in this *single-ion magnet* (SIM) is determined by fluctuations of the local magnetic field due to the impurity Ho$^{3+}$ ions, as evidenced by the fact that the measured relaxation rates for the LiY$_{0.998}$Ho$_{0.002}$F$_4$ sample are about three orders of magnitude larger than for pure LiYF$_4$ [10]. The spin dynamics at the anti-crossings of the hyperfine sublevels of the ground electronic doublet of Ho$^{3+}$ ions produce striking effects at low temperatures, comparable to those observed in SMM. Large peaks (dips) were observed in the magnetic field dependence of the ac-susceptibility [11] and the fluorine nuclear relaxation rate [10] at temperatures in the range 1.7-4 K. The ac-susceptibility results were subsequently described by a microscopic theory [12], while the NMR measurements were analyzed using a phenomenological model [4,5].

It is highly desirable to go beyond the phenomenological model presented in Ref. 10 and carry out microscopic calculations describing the $^{19}$F spin-lattice relaxation measurements, as was done for the ac-susceptibility in Ref. 14. This requires one to take into account the spin-diffusion processes induced by cross-relaxation between the nuclei, which transport the spin temperature across the sample from the nuclei which are relaxed directly by the low concentration of paramagnetic ions [13-19]. As a result, both non-exponential and exponential time evolution of the nuclear magnetization may be observed in different time domains, depending on the specific parameters of the system under consideration.

In the present work, microscopic calculations of the $^{19}$F nuclear relaxation are carried out by (a) making use of the earlier developed theory of, and calculations for, electronic relaxation in holmium-doped LiYF$_4$ [12] and (b) by taking into account the consequences of the spin-diffusion theory [18,19]. The experimental data obtained earlier [10] are reanalyzed, and additional measurements have been performed to observe the predicted variation of the fluorine spin-lattice relaxation time at the anti-crossings of electron-nuclear sublevels of Ho$^{3+}$ ions with the magnetic field component transverse to the symmetry axis of a crystal. Analysis of the relaxation rates for magnetic fields far from and in between the crossing points allowed us to determine the parameters of nuclear spin-diffusion processes. The previously obtained parameters of Ho$^{3+}$ cross-relaxation (CR) processes [12] have been corrected to describe the behavior of fluorine nuclear relaxation as well as of holmium spin dynamics at the anti-crossings. In all cases we find our calculations in good agreement with experiments, where the only fitting parameters are associated with $^{19}$F spin diffusion and holmium CR rates.

This investigation of electronic spin dynamics in the highly diluted paramagnet LiYF$_4$:Ho$^{3+}$ should be useful for understanding peculiarities not only of the nuclear relaxation rates at avoided level crossings, but also of the measured frequency and temperature dependences of the ac-susceptibility in diluted ferromagnets LiHo$_c$Y$_{1-c}$F$_4$, for which the anomalous frequency dependences at intermediate and low temperatures remain an open question [20,21].

The paper is arranged as follows. In Section 2 we present general expressions for nuclear relaxation rates in highly diluted paramagnetic crystals. Experimental data and general tests of the model are described in Section 3, while Section 4 contains the detailed results of simulations of magnetic field and temperature dependences of $^{19}$F nuclear relaxation, which are further compared with the measured data. In Section 5 we conclude and summarize the main results.

## 2 Theoretical background

We present here a brief derivation of the expression for the nuclear spin-lattice relaxation rate which is used below in the analysis of the experimental data. In an insulating crystal containing paramagnetic ions, the spin-lattice relaxation time $T_1$ for the host lattice nuclei with spin $I = 1/2$ is determined by transition probabilities between the nuclear states $|+>$ and $|->$ (the eigenfunctions of the nuclear spin projection on the local magnetic field $\boldsymbol{B}_{loc}$ at a nucleus) induced by fluctuations of the local field.

The transition probability is determined through the spectral densities of correlation functions for fluctuations of the transverse components of the local magnetic field [2]:

$$w = \frac{\gamma^2}{4\hbar^2} \{ J_{\Delta B_{loc,x'} \Delta B_{loc,x'}}(\omega) + J_{\Delta B_{loc,y'} \Delta B_{loc,y'}}(\omega) \}, \tag{1}$$

where γ is the nuclear gyromagnetic ratio, $\omega = \gamma B_{loc}$ is the local Larmor frequency, the z' axis of the local frame is parallel to $\boldsymbol{B}_{loc}$, and $J_{AB}(\omega)$ is the spectral density of the correlation function $<A(t)B(0)>$. Angular brackets denote an average taken with the equilibrium density matrix $\rho$ of a single paramagnetic ion. The local magnetic field at the host nucleus equals the sum of the external dc magnetic field $\boldsymbol{B}$ and the dipole fields from impurity paramagnetic ions with magnetic moments $\boldsymbol{m}$ ($\boldsymbol{m} = -g_J \mu_B \boldsymbol{J}$ for rare earth ions, $g_J$ is the Landé-factor of the ground multiplet, $\mu_B$ is the Bohr magneton, and $\boldsymbol{J}$ is the angular momentum). We adopt the usual assumption that the sample is divided up into "regions of influence" where only one of paramagnetic ions is important in determining the nuclear relaxation. The region of influence is assumed to be a sphere centered on a paramagnetic ion, and of radius $R$ equal to the average separation of paramagnetic ions ($R = (4\pi N/3)^{-1/3}$, where $N$ is the number of paramagnetic ions per unit volume). Because only nuclei at large enough distances $r > r_0$ from the paramagnetic ion contribute to the NMR signals, spin-transfer effects are neglected (within the region of critical radius $r_0$ nuclei have Larmor frequencies which are shifted outside the resonance line at the frequency $\omega_0 = \gamma B$).

Fluctuations of the local magnetic field can be easily written through the fluctuations of the magnetic moment $\Delta \boldsymbol{m} = \boldsymbol{m} - \langle \boldsymbol{m}\rangle$ of a paramagnetic ion. We obtain from Eq. (1)

$$w = C(\boldsymbol{r}/r,\boldsymbol{B},T)/2r^6, \qquad (2)$$

where

$$C(\boldsymbol{r}/r,\boldsymbol{B},T) = \frac{\gamma^2}{2}\{J_{\Delta m \Delta m}(\omega) + 3J_{\Delta m_r \Delta m r}(\omega) - J_{\Delta(m_B - 3m_r e_{rB})\Delta(m_B - 3m_r e_{rB})}(\omega)\}, \qquad (3)$$

and $\Delta m_r = \Delta \boldsymbol{m}\cdot \boldsymbol{r}/r$, $\Delta m_B = \Delta \boldsymbol{m}\cdot \boldsymbol{B}_{loc}/B_{loc}$, and $e_{rB} = \boldsymbol{r}\cdot \boldsymbol{B}_{loc}/rB_{loc}$. The spectral densities of the single-ion correlation functions for the electronic magnetic moment can be expressed through the dynamic susceptibility $\chi(\omega)$ of a paramagnetic ion at the temperature $T$, using the fluctuation-dissipation theorem ($k_B$ is the Boltzmann constant):

$$J_{\Delta m_\alpha \Delta m_\beta}(\omega) = \frac{2k_B T}{\omega}\mathrm{Im}\chi_{\alpha\beta}(\omega) \qquad (\hbar\omega \ll k_B T). \qquad (4)$$

The dynamic susceptibility has the form [12,22]

$$\chi_{\alpha\beta}(\omega) = \chi^0_{\alpha\beta} - i\omega\sum_{nk}\Delta m_{\alpha,nn}(i\omega\boldsymbol{I}+\boldsymbol{W})^{-1}_{nk}\Delta m_{\beta,kk}\rho_{kk}/k_B T +$$
$$\sum_{n,k\neq n}m_{\alpha,nk}m_{\beta,kn}(\rho_{kk}-\rho_{nn})\left[\frac{1}{\hbar(\omega_{nk}-\omega-i\gamma_{nk})}-\frac{1}{\hbar\omega_{nk}}\right], \qquad (5)$$

where $\chi^0_{\alpha\beta}$ is the static susceptibility, $\rho_{kk}$, $\boldsymbol{m}_{kn}$ and $\Delta \boldsymbol{m}_{kk}$ are matrix elements of the equilibrium density matrix, the magnetic moment and its fluctuation, respectively, in the basis of eigenvectors of the single ion Hamiltonian with the eigenvalues $E_k$, $2\gamma_{nk}$ is the homogeneous width of the transition $n \leftrightarrow k$ at the frequency $\omega_{nk} = (E_n - E_k)/\hbar$, and $\boldsymbol{W}$ is the relaxation matrix. A non-diagonal element $W_{nk}$ of the relaxation matrix is the probability of the transition $k \to n$ which may be induced by the electron-phonon interaction and different interactions (magnetic dipole-dipole, in particular) between paramagnetic ions, $W_{nn} = -\sum_k W_{kn}$. The explicit expressions for the transition probabilities were presented in Ref. 14 where the renormalization of the electron-phonon coupling due to the phonon bottleneck effect and contributions due to energy exchange between paramagnetic ions were accounted for. In particular, the two-ion CR processes are described by the transition probabilities

$$W_{nm}^{CR} = \sum_{lp}(W_{np,lm}^{CR}\rho_{pp} + W_{nm,lp}^{CR}\rho_{pp} - W_{pn,lm}^{CR}\rho_{nn}) \quad (n \neq m), \tag{6}$$

where $W_{np,lm}^{CR}$ is the probability for the simultaneous transitions $m \to l$ of one paramagnetic ion, and $p \to n$ of another ion.

Using Eq. (4), we obtain the corresponding spectral density of fluctuations of the magnetic moment of a paramagnetic ion

$$J_{\Delta m_\alpha,\Delta m_\beta}(\omega) = -2\,real\{\sum_{nk}\Delta m_{\alpha,nn}(i\omega\mathbf{1}+\mathbf{W})_{nk}^{-1}\Delta m_{\beta,kk}\rho_{kk}\}$$
$$+ \frac{4k_BT}{\hbar}\sum_{nk}\frac{m_{\alpha,nk}m_{\beta,kn}(\rho_{kk}-\rho_{nn})\omega_{nk}\gamma_{nk}}{(\omega_{nk}^2-\omega^2)^2+4\omega^2\gamma_{nk}^2} \quad . \tag{7}$$

In order to connect our microscopic approach to models currently used to describe nuclear relaxation in SMM, we note that the first term in Eq. (7) corresponds to the quasi-elastic relaxation contribution, while the second term is caused by direct energy exchange between paramagnetic ions and host nuclei, and corresponds to the inelastic relaxation included in the phenomenological model described in Refs. 4,5. This may be the dominant contribution to relaxation near anti-crossings at low temperatures.

Having obtained a microscopic expression for the spectral fluctuation density, which can be calculated using known parameters for the titled system [12], we now consider the effect of spin diffusion on the $^{19}$F nuclear magnetization recovery curve. Since in the presence of spin diffusion the relaxation recovery curves are typically non-exponential, it is not possible in general to define a $T_1$ parameter. Even when the recovery is exponential, the measured $T_1$ is a complicated function of transition probabilities induced by the direct interaction of a given nucleus with the paramagnetic ion at the distance *r* (see Eq. (2)) and of a propagation (diffusion) rate of non-equilibrium spin polarization within the ensemble of nuclei. An explicit form of this function depends on the spin-diffusion rate and on the concentration and electronic relaxation rates of impurity ions.

After the macroscopic nuclear magnetization *M* is saturated, it recovers towards equilibrium ($M_0$ is the equilibrium magnetization) due to processes of direct relaxation:

$$M(t) - M_0 = [M(0) - M_0] \frac{\sum \exp[-C(\mathbf{r}/r, \mathbf{B}, T) t / r^6]}{\sum 1}. \tag{8}$$

Here the sums are taken over nuclei at the distances $r_0 \leq r \leq R$ from the paramagnetic ion. Results of direct calculations of these sums (in particular, in the titled SIM) for large enough *t* (see below) are well approximated by the expression

$$M(t) - M_0 \sim \exp[-(\pi C_{Av}(\mathbf{B}, T) t / R^6)^{1/2}], \tag{9}$$

where $C(\mathbf{B}, T)_{Av} = \sum C(\mathbf{r}/r, \mathbf{B}, T) / \sum 1$ ($C_{Av}$ does not depend on *R* when *R* is much larger than the lattice constants), and $C_{Av} t / r_0^6 > 1$. This form of the evolution of the non-equilibrium nuclear magnetization shows that it is possible to introduce the isotropic direct spin-lattice relaxation time

$$(1/T_1)_r = C(\mathbf{B}, T)_{Av} / r^6 \tag{10}$$

that brings about the same functional time dependence of the nuclear magnetization in the framework of the continuum approximation [23]. The stretched-exponential recovery of nuclear magnetization has been observed in many systems (in particular, in $NH_4HSO_4:Cr^{3+}$ [14] and $CaF_2:Eu^{3+}$ [24]). However, the $t^{1/2}$-region in the magnetization recovery may be absent in the case of rapid spin diffusion when a magnetization flux reaches distant nuclei before they contribute to the bulk relaxation due to the direct interaction with the impurity paramagnetic ion. In any case at long times the functional form of the time-dependent magnetization is determined by the spin-diffusion.

In the continuum approximation, and neglecting the anisotropy of the diffusion tensor, the nuclear magnetization density $M(\boldsymbol{r},t)$ satisfies the equation [13,18]

$$\frac{\partial M(\boldsymbol{r},t)}{\partial t} = -\sum_{k}(1/T_1)_{|\boldsymbol{r}-\boldsymbol{r}_k|}(M(\boldsymbol{r},t)-M_0) + D(r)\Delta M(\boldsymbol{r},t), \tag{11}$$

where the sum is taken over all paramagnetic ions, and the position dependent diffusion is described by $D(r<\delta)=0$ and $D(r>\delta)=D$, with $D$ being the spin-diffusion constant. The radius $\delta$ of the diffusion barrier (assumed isotropic) equals the distance from the paramagnetic ion for which the difference between magnetic fields produced by the ion at neighboring nuclei is approximately equal to the local nuclear dipolar field, and is proportional to $<m>^{1/4}$ [15]. For a small concentration of paramagnetic ions, the asymptotic time dependence of the bulk nuclear magnetization $M(t) = \frac{1}{V}\int M(\boldsymbol{r},t)d\boldsymbol{r}$ ($V$ is the sample volume) takes the form [18,19]

$$M(t) - M_0 = (M(0) - M_0)\exp(-t/T_1 - (t/T_s)^{1/2}). \tag{12}$$

Here

$$1/T_1 = \frac{3D\lambda}{R^3}, \quad \frac{1}{T_s} = \frac{12}{\pi}\left(\frac{\lambda}{R}\right)^3\frac{1}{T_1} = \frac{4}{9\pi}\left(\frac{\tau_D}{T_1}\right)^3\frac{1}{T_1}, \tag{13}$$

$\tau_D = R^2/D$ is the diffusion time necessary for the magnetization flux to propagate at distance $R$ from a paramagnetic ion, and $\lambda$ is the characteristic distance from the paramagnetic ion for which spin-diffusion begins to dominate over direct relaxation. The value of $\lambda$ (see Ref. [16])

$$\lambda = b\frac{2x\mathrm{I}_{3/4}(x)}{\mathrm{I}_{1/4}(x) + 2x\mathrm{I}_{5/4}(x)}, \tag{14}$$

where $\mathrm{I}_p(x)$ is a Bessel function of imaginary argument and $x = (C_{Av}/D)^{1/2}(1/2\delta^2)$, depends on the relative values of the diffusion radius $b = \frac{\pi}{4\sqrt{2}[\Gamma(\frac{5}{4})]^2}(C_{Av}/D)^{1/4}$ and the diffusion barrier $\delta$ ($\Gamma(x)$ is the Euler Gamma-function) [14-18]. In the extreme case of diffusion limited relaxation, $b>>\delta$, $\lambda = b$, and $1/T_1 \sim C_{Av}^{1/4}$. In the opposite case of rapid spin-diffusion, $b<<\delta$, $\lambda = C_{Av}/3D\delta^3$ and $1/T_1 \sim C_{Av}$.

As discussed in Ref. 18, Eqs. (12) and (13) are valid if $N\lambda^3 << 1$, and the term $(t/T_s)^{1/2} \sim (t/T_1)^{1/2}(N\lambda^3)^{1/2}$ may be neglected. However, as the concentration of paramagnetic ions becomes larger, we enter the diffusion-limited regime, and this term may have a significant effect on the observed nuclear magnetization at times $t \sim T_1$. In any case, with the increasing time, the kinetics of the magnetization recovery becomes slower than simple exponential time dependence [19].

## 3 Experimental

In this section we describe experimental results, and test some general predictions based on our calculations for the spectral densities of fluctuations of the $Ho^{3+}$ magnetic moment and on the inclusion of spin diffusion effects to determine the nuclear relaxation rate. Detailed simulation results will be discussed in the next section.

First of all, we remind briefly the spectral properties of impurity $Ho^{3+}$ ions in $LiYF_4$ (the crystal lattice has the $C_{4h}^6$ space group, the lattice constants are given in Ref. [25]) which substitute for $Y^{3+}$ ions in sites with $S_4$ symmetry. There is only one holmium isotope, $^{165}Ho$, with nuclear spin $I = 7/2$. The free ion ground state, the $^5I_8$ multiplet, is split by the tetragonal crystal field into 13 sublevels with a $\Gamma_{34}$ doublet ground state and the first excited $\Gamma_2$ singlet at 6.85 cm$^{-1}$ higher energy (here $\Gamma$ is the corresponding irreducible representation of the $S_4$ point symmetry group) [26,27]. The electronic and nuclear states are mixed mainly by the magnetic hyperfine interaction $H_{hf}=A\mathbf{J}\cdot\mathbf{I}$ ($A = 0.795$ GHz [26]). The calculated hyperfine structure of the ground electronic doublet (the corresponding g-factors equal $g_\parallel$=13.3, $g_\perp$=0 [26]) in the external magnetic field $B(\sin\theta\cos\varphi, \sin\theta\sin\varphi, \cos\theta)$ with the orientation defined by angles $\theta$ = 27.5$^0$ and $\varphi$ = 45$^0$ relative to the crystallographic axes ($z\|c$, $x\|a$) is shown in Fig. 1. The spectrum, which is nearly independent of the angle $\varphi$, consists of two almost equidistant groups of electron-nuclear sublevels with positive and negative slopes. The sublevels with different slopes cross at magnetic fields $B \cong |m'+m|\Delta B/\cos\theta$, where $m$ and $m'$ are the quantum numbers for the z-components of the nuclear spin for the crossing levels, and, to first approximation in the hyperfine constant $A$, $\Delta B = A/2g_J\mu_B$. There are odd and even crossing points corresponding to magnetic fields $B = (2n+1)\Delta B/\cos\theta$ (for $|m'-m| = \Delta m = 2k$) and $B = 2n\Delta B/\cos\theta$ (for $|m'-m| = \Delta m = 2k+1$), ($k, n = 0, 1, 2, 3$), respectively. The magnetic transverse hyperfine interaction opens gaps at the odd crossing points with the maximum value of 0.35 GHz for $\Delta m = 2$ (see Fig. 1) due to mixing of the ground doublet with the excited singlets. Additional gaps of comparable magnitude at the odd crossing points for $\Delta m = 0$ are induced by random crystal fields [26]. Within the context of the model presented here, gaps at the even crossings can only be induced by a transverse external field component $B\sin\theta$; for $\theta \sim 30^0$ the largest gaps created by the transverse field component occur at the $\Delta m = 1$ crossings and are comparable to the gaps at the odd anti-crossings (see Fig. 1).

Measurements of fluorine NMR spectra and of the nuclear magnetization recovery following a saturating sequence of $\pi/2$ pulses were performed on a single crystal with holmium concentration of 0.127 at. wt. %, corresponding to a Ho$^{3+}$ number density of $N$ = 1.8·10$^{-2}$ nm$^{-3}$, or about 3100 fluorine nuclei for each impurity ion. The concentration was determined from a comparison of the static magnetic susceptibility, measured with a SQUID magnetometer at temperature $T = 2$ K, with the calculated susceptibility of the single Ho$^{3+}$ ion.

The magnetic field dependence of the relaxation rate was studied at temperatures 1.7 – 4.2 K in the range of magnetic fields from 0.1 to 0.2 T for different angles between the field and the *c*-axis (the data for $\theta \sim 30^0$ were presented earlier in Ref. [10]). This magnetic field range includes four groups of anti-crossing points (N4-N7, see Fig. 1). The recovery of the nuclear magnetization at low temperatures is nearly exponential and can be described by a single parameter $1/T_1$ (that is, in the rapid diffusion limit). The measured field dependence of $1/T_1$ contains a background that decreases weakly with the increasing field, and exhibits peaks in the vicinity of the crossing points. These peaks are quite sharp (full width at half height of approximately 2 mT) at the lowest temperature of 1.7 K (see Fig. 2), but lose intensity relative to the background at higher temperatures and disappear for $T > 4$ K. Enhancement of the nuclear relaxation rate at the anti-crossing points may be caused by: (1) large changes of the relaxation matrix ***W*** due to the cross-relaxation or the mixing of wave functions of the crossing energy levels, and (2) quasi-resonant energy exchange between the paramagnetic ions and the host nuclei. When a difference between the transition frequency $\omega_{nk}$ and the nuclear Larmor frequency $\omega$ becomes comparable to the transition width $\gamma_{nk}$ (see Eq. (7)), the efficiency of the latter mechanism is determined by the matrix elements of the electronic magnetic moment connecting crossing electron-nuclear states, and also depends on the mixing of wave functions of these energy levels with positive and negative slopes. According to calculations, it is this latter mechanism which dominates the response at anti-crossings (see Fig. 2).

For large enough angle $\theta$ (in particular, for $\theta = 27.5^0$, see Figs. 1 and 2), the largest gaps at all the anti-crossings N4-N7 have comparable values (these values characterize the degree of mixing of the corresponding wave functions), and so the corresponding peaks in the nuclear relaxation have comparable heights. However, the extreme peak N7 has lower intensity, and requires a special discussion presented in Section 4. With decreasing angle $\theta$, calculations predict that the gaps at the odd anti-crossings remain almost unchanged, while the gaps at the even anti-crossings go to zero. Therefore, for $\theta = 0$, the even numbered peaks in $1/T_1$ should essentially disappear because the direct energy exchange between the Ho$^{3+}$ ions and fluorine nuclei becomes ineffective (the dominant role of the energy exchange is clearly evident when comparing the curves (1,2) with the curve (3) in Fig. 2). A similar result was predicted from single-ion calculations for the ac-susceptibility [12]. However measurements of the ac-susceptibility on dilute samples (with holmium concentration $x = 0.001$) in fact did show sharp changes at the even numbered crossings for $\theta = 0$, as well as small structures midway between the single-ion level crossings [11]. These discrepancies were successfully accounted for by including two-ion energy exchange (CR) processes in the microscopic model [12].

In order to experimentally determine the effect of the gap on the nuclear relaxation peaks and whether CR is also important in nuclear relaxation processes, we studied in the present work the angular variation of the peak heights in $1/T_1$ at different anti-crossings. Measurements of $T_1$ were carried out at $T = 1.7$ K for angles $\theta = 27.5^0$, $20^0$, $16^0$, $11^0$ and $2^0$, where the angles between the magnetic field and the *c*-axis were determined by a comparison of the measured and calculated magnetic field values corresponding to peaks in the relaxation rates; the observed critical fields are larger than the calculated fields by a factor $1/\cos\theta$. The observed shapes and heights of the peaks at the anti-crossing points do not change down to an angle $\theta = 16^0$, and then the intensities of peaks N6 and N7 begin to diminish (see Fig. 3) while the peak N5 remains unchanged except for a slight increase in the width. The observed minimum in the angular dependence of the peak N6 unambiguously demonstrates that the CR processes are less important for the nuclear spin-lattice relaxation than for ac-susceptibility. However, the residual peak observed at N6 for small angles results from CR processes.

We now apply the general considerations of spin diffusion effects as discussed in Section 2 to previously obtained data for the temperature-dependent nuclear magnetization recovery at fixed magnetic fields [10]. These measurements were carried out in the range of temperatures from 4 to 50 K and in magnetic fields $B = 0.1335$ (in between the anti-crossing points), 0.332 and 0.751 T (outside the region of crossings) with magnetic fields oriented at $27.5^0$ relative to the *c*-axis. Several plots of the time evolution of the nuclear magnetization recovery are shown in Fig. 4, and one immediately notices that the recovery cannot be approximated by a single exponential function of time. At very short times there is no gradient of the magnetization density, and direct relaxation results in the stretched exponential time dependence (see Eq. (9)). Then for $t > b^2/D$ spin-diffusion changes the character of the time evolution, which can be approximated by a product of exponential and stretched exponential functions as given in Eq. (12). We emphasize that the decay curves in the range 5 – 20 K do not follow simple exponential behavior even at long times, a result which has been predicted in dilute paramagnetic systems [18] but, to our knowledge, never observed. However, due to the fairly large measurement error for small values of the magnetization, it is difficult to obtain unique values for the parameters $T_1$ and $T_s$ from these individual plots. We have therefore introduced the additional constraint $1/T_s = K(1/T_1)^4$ where the value of the parameter $K$ is assumed to be independent of temperature and magnetic field strength, as given in Eq. (13).

The relaxation rates $1/T_1$ and $1/T_s$ obtained from the fitting of the measured recovery plots by Eq. (12) with the parameter $K = 100$ (msec)$^3$ are presented in Fig. 5. The relaxation rates have maximum values at temperatures $T_{\text{Max}}$ which shift to higher temperatures with the increasing magnetic field. The values of $k_B T_{\text{Max}}$ are close to the energy of the first excited singlet in the spectrum of Ho$^{3+}$ ions, suggesting that these maxima result from an exponentially increasing relaxation rate via the first excited singlet. The simulations presented in the next section confirm this interpretation.

## 4 Simulations of the $^{19}$F relaxation rates

The nuclear relaxation rates were simulated using the explicit form (Eq. (7)) of the spectral densities of the correlation functions for the magnetic moment of the Ho$^{3+}$ ion. We worked with the single-ion Hamiltonian operating in the space of the 136 electron – nuclear states of the ground multiplet $^5I_8$ and containing the crystal field, Zeeman and magnetic hyperfine interactions. The elements of the relaxation matrix $W$ for impurity Ho$^{3+}$ ions in the LiYF$_4$ crystal have already been calculated in the study of the ac-susceptibility [12]. The probabilities of one-phonon transitions between the 64 electron-nuclear sublevels of lower crystal field energy levels of the Ho$^{3+}$ ion with excitation energies below 300 K were computed at fixed temperatures and external magnetic fields using wave functions obtained from numerical diagonalization of the single-ion Hamiltonian and electron-phonon coupling constants presented in Ref. 12. The effects due to low-symmetry random crystal fields were accounted for by introducing the operator $H_{RCF} = \alpha(B_2^2 O_2^2 + B_2^{-2}\Omega_2^2)$ (where $O_2^2$ and $\Omega_2^2$ are the Stevens operators and α = - 1/450 is the corresponding reduced matrix element for the $^5I_8$ multiplet) with parameters $B_2^2$ =0.4525, $B_2^{-2}$ = -0.4752 cm$^{-1}$, which have been already used earlier [26] to describe the anti-crossings with Δ$m$ = 0 observed in the EPR spectra of a LiYF$_4$:Ho$^{3+}$ (0.1%) crystal.

Calculations of relaxation rates of individual fluorine nuclei from Eq. (3) show that the maximal values of $C(\boldsymbol{r}/r,\boldsymbol{B},T)/r^6$ for different distances $r$ from the Ho$^{3+}$ ion are described by the function $C_{\text{Max}}(B,\theta,T)$ that depends only on temperature, magnetic field strength and the angle between the magnetic field and the $c$-axis. The functions $C_{\text{Av}}(\boldsymbol{B},T)$ and $C_{\text{Max}}(B,\theta,T)$ are related via $C_{Av}(\boldsymbol{B},T) = kC_{Max}(B,\theta,T)$, where the coefficient $k = 0.438$ is found to be independent of the magnetic field and temperature. Relaxation rates of fluorine nuclei with the Larmor frequencies outside the range $\gamma B \pm 2\pi\Delta\nu$, where $\Delta\nu = 20$ kHz is the NMR line half width [10], have not been taken into account. The simulated temperature dependences of $C_{Av}$ in the magnetic fields used in this study are presented in the insert in Fig. 5. The temperature dependences of $C_{Av}$ and $1/T_s$ are proportional to each other, as expected from the definition Eq. (13) of the rate $1/T_s$ for the case of the diffusion limited relaxation, although the coefficient of proportionality in fact depends on the magnetic field strength. The positions of the maxima $T_{Max}$ in the calculated and measured temperature dependences of the relaxation rates are found to coincide. It should be stressed that the direct relaxation rates at $T_{Max}$ are two orders of magnitude larger than those at low temperatures ($T = 1.7 - 2$ K).

At low temperatures (in particular, at $T = 1.7$ K) and magnetic fields ($B \sim 0.1$-$0.2$ T), when the spin diffusion processes dominate, the value of the radius $\delta$ of the diffusion barrier, estimated from the relation $\dfrac{1}{T_1} = \dfrac{C_{Av}}{(R\delta)^3}$ by making use of the measured relaxation rates and the calculated function $C_{Av}(\boldsymbol{B},T)$, is close to 0.9 nm; the average radius $R$ of the spherical volume per paramagnetic ion as determined from the measured concentration of impurity ions equals 2.4 nm. The relaxation rate $1/T_1$ is proportional to $C_{Av}$ only in a case of rapid spin-diffusion (if $\delta \gg b$). Estimates of the diffusion constant $D$ (see below), show that in this range of temperatures and fields the diffusion radius $b \sim 0.5$-$0.6$ nm, which is in fact less than $\delta$ but has a comparable value. Therefore, in order to extract the experimental time evolution of the nuclear magnetization from the computed direct relaxation rates, we used the analytical expression Eq. (14) for the diffusion length $\lambda$, with $\delta$ and $D$ as fitting parameters.

The simulated field dependence of the relaxation rate $1/T_1$ in the region of anti-crossings at $T = 1.7$ K is presented in Fig. 2. The monotonic decrease of the background relaxation rate with increasing magnetic field is well described by simulations if we choose the (field independent) radius $\delta = 0.86$ nm and $D = 400$ nm$^2$/sec. This value for the diffusion constant is reasonable when compared, for example, to the directly measured spin-diffusion coefficients in cubic CaF$_2$ crystal 530±30 nm$^2$/sec and 710±50 nm$^2$/sec along [111] and [001] directions, respectively [28], and to the recently measured diffusion constant $D = 620\pm70$ nm$^2$/sec in the same crystal by making use of magnetic resonance force microscopy [29]. Redfield and Yu derived the following expression for the averaged diffusion constant appropriate for the problem under consideration in the present work [30]:

$$D = \frac{\sqrt{\pi}}{60} \gamma^2 \hbar \left( \sum_j r_{jk}^{-4} \right) \left( < \sum_j (1 - 3\cos^2 \theta_{jk})^2 r_{jk}^{-6} > \right)^{-1/2}. \quad (15)$$

Here $r_{jk}$ is the vector connecting two nuclei, and $\theta_{jk}$ is the angle between $r_{jk}$ and the dc magnetic field. From calculations of the corresponding sums for the LiYF$_4$ crystal lattice, we obtained the value of the diffusion constant $D = 422$ nm$^2$/sec that matches almost exactly our empirical result (the more recent theoretical model [31] brings about the similar result). We note also that the predicted small anisotropy of the diffusion tensor [30] should not significantly affect results of calculations of the nuclear relaxation rates.

In Ref. 12 it was found that the calculations must include the phonon bottleneck effect in order to describe the ac-susceptibility data at low temperatures ($T \leq 4$ K), as it determines the ratio of peak intensities at the anti-crossings to the intensity of the background. Following those results we have used phonon lifetimes ~1 μs and ~0.1 μs for resonant phonons with frequencies corresponding to transitions between the electron-nuclear sublevels of the ground doublet, and between the first excited singlet and the ground doublet of a Ho$^{3+}$ ion, respectively [12].

As it is seen in Fig. 3, where the field dependences of the real and imaginary parts of the low-frequency ac-susceptibility measured with the laser-polarimetric technique [32] are compared with the field dependences of the nuclear relaxation rate, the peaks in $1/T_1$ at the anti-crossing points correlate with changes of electronic relaxation rates of $Ho^{3+}$ ions (see also [11,12]). The strong enhancement of the calculated nuclear relaxation rate at the anti-crossing points is caused by cross-relaxation processes in the subsystem of impurity paramagnetic ions and direct energy exchange processes, as given by the last (inelastic) term in the spectral density Eq. (7), where we have used the same value of the homogeneous width $\gamma_{nk} = 2\pi \cdot 1.4 \cdot 10^8$ sec$^{-1}$ for all transitions. This value agrees with the measured line widths in the submillimeter EPR spectra of the $LiYF_4$:$Ho^{3+}$ (0.1%) crystal [26]. As can be seen in Fig. 2, the calculated height and width of the peak at the highest field anti-crossing N7 are very sensitive to the choice of parameters characterizing the random crystal field. Random crystal fields also produce broadening of the peak N5, as the $\Delta m = 0$ and $\Delta m = 2$ anti-crossings are slightly shifted in field relative to each other (see Fig. 1).

From low temperature measurements of the nuclear relaxation at the anti-crossing points, we obtained additional information that allowed us to revise values of parameters used earlier in Ref. 12 in the definition of the CR rates. The transition probabilities in Eq. (6) were presented in the following form

$$W_{np,lm}^{CR} = d^2 \{ g_{11}^{CR}(\omega_{pn} - \omega_{lm}) k_{11}(|J_{x,np} J_{x,lm}|^2 + |J_{y,np} J_{y,lm}|^2) +$$
$$g_{12}^{CR}(\omega_{pn} - \omega_{lm}) k_{12}(J_{x,np} J_{x,lm} J_{y,pn} J_{y,ml} + J_{y,np} J_{y,lm} J_{x,pn} J_{x,ml}) +$$
$$g_{13}^{CR}(\omega_{pn} - \omega_{lm}) k_{13}[J_{z,np} J_{z,lm}(J_{x,pn} J_{x,ml} + J_{y,pn} J_{y,ml}) + (J_{x,np} J_{x,lm} + J_{y,np} J_{y,lm}) J_{z,pn} J_{z,ml}] +$$
$$g_{33}^{CR}(\omega_{pn} - \omega_{lm}) k_{33} |J_{z,np} J_{z,lm}|^2 + g_{66}^{CR}(\omega_{pn} - \omega_{lm}) k_{66} |J_{x,np} J_{y,lm} + J_{y,np} J_{x,lm}|^2 +$$
$$g_{44}^{CR}(\omega_{pn} - \omega_{lm}) k_{44}(|J_{x,np} J_{z,lm} + J_{z,np} J_{x,lm}|^2 + |J_{y,np} J_{z,lm} + J_{z,np} J_{y,lm}|^2) \} +$$
$$g_{\varepsilon}^{CR}(\omega_{pn} - \omega_{lm}) \varepsilon^2 (|O_{2,np}^2 O_{2,lm}^2|^2 + |\Omega_{2,np}^2 \Omega_{2,lm}^2|^2),$$

(16)

where $g^{CR}(\omega)$ are the cross-relaxation line shapes (Gaussians with a dispersion of 100-200 MHz as in Ref. 12). Note that we have included an additional contribution (the last term in Eq. (16)) corresponding to the phonon exchange between the $Ho^{3+}$ ions. Values of parameters $d = 0.773$ and $\varepsilon = 0.0154$ in units of $10^8$ sec$^{-1}$, $k_{11} = 0.798$, $k_{12} = -0.675$, $k_{13} = -0.00385$, $k_{33} = 0.0079$, $k_{44} = 0.13$, $k_{66} = 0.586$ were obtained from the simultaneous fitting of field dependences of the ac-susceptibility and the fluorine nuclear relaxation rates.

The theory predicts important changes of the peaks in the $^{19}$F spin lattice relaxation rate at the even anti-crossings with the diminishing transverse component of the external magnetic field. When the direction of the field approaches the *c*-axis, shapes and intensities of all the peaks at the anti-crossings in the calculated field dependences of the relaxation rate $1/T_1$ remain almost unchanged until the value of the angle $\theta$ is approximately 11 degrees, below which the intensities of the peaks at even anti-crossings begin to diminish. In Fig. 3 the calculated inelastic and cross-relaxation contributions to the nuclear relaxation rates are compared with the experimental data for three values of the angle $\theta$. The specific behavior of the peak N6 in the calculated field dependences correlates qualitatively with the results of measurements. Some differences may be caused either by a slight misorientation, or random deformations of the crystal lattice. Finally, we note that according to measurements, the intensity of the peak N7 also diminishes slightly with the decreasing angle $\theta$. This effect can be related to high sensitivity of this peak to the homogeneous widths of transitions between the electron-nuclear sublevels of the ground doublet of the Ho$^{3+}$ ions, which may change with transverse magnetic field component; this effect would not be evident in our calculations, as we have assumed that the homogeneous width is the same for all transitions.

The temperature dependences of $1/T_1$ at different magnetic fields were computed by making use of Eq. (14) for the diffusion length $\lambda(\boldsymbol{B},T)$. Results of calculations are compared with the experimental data in Fig. 5. The diffusion barrier $\delta$ is expected to increase with the magnetic field strength and to decrease with temperature. To account for these variations, we introduce the average absolute value $<m>$ of the effective magnetic moment of Ho$^{3+}$ ions in order to determine $\delta(\boldsymbol{B},T)$ via

$$\delta(\boldsymbol{B},T) \sim <m>^{1/4} \sim \left( \sum r^3 \mid B - \omega(\boldsymbol{r},\boldsymbol{B},T)/\gamma \mid / \sum 1 \right)^{1/4} \qquad (17)$$

where the sum is taken over all fluorine nuclei inside the influence sphere of radius $R$ with Larmor frequencies $\omega(\boldsymbol{r},\boldsymbol{B},T)$. Starting from the value of $\delta$ =0.86 nm used above in the simulation of the relaxation rate in the magnetic field 0.1335 T at $T$ = 1.7 K, we obtained $\delta$ =1.01 nm and 1.06 nm in the fields 0.332 T and 0.751 T, respectively, at the same temperature (all values refer to magnetic fields declined from the $c$-axis by 27.5$^0$). At higher temperatures, values of $\delta$ were calculated according to Eq. (17).

It is seen in Fig. 5 that the simulated relaxation rates $1/T_1$ yield adequate fits to the experimental data. However, the rates $1/T_s$ of the stretched exponential relaxation calculated according to Eq. (13), while qualitatively describing the data, are larger than the measured rates by nearly a factor of four. This discrepancy shows that a more elaborate model of spin diffusion is required for a precise description of the data. Nonetheless, the theory presented in Section 2 gives a semi-quantitative description of the nuclear relaxation. Moreover, it predicts the occurrence of a stretched-exponential recovery of the $^{19}$F nuclear magnetization in the system LiYF$_4$:Ho$^{3+}$ with the holmium concentration of about 0.1 % which has indeed been observed experimentally.

## 5 Conclusion

We described satisfactorily the measured temperature and magnetic field dependences of the $^{19}$F nuclear relaxation rates in LiYF$_4$:Ho$^{3+}$ crystals using the microscopic model of electron-phonon interaction with previously determined physical parameters and only introducing fitting parameters to characterize the electronic CR and nuclear spin diffusion processes. The self-consistent interpretation of field dependences of the low frequency ac-susceptibility and the $^{19}$F nuclear relaxation rate at liquid helium temperatures is achieved by making use of the single set of parameters that determine the transition probabilities induced by magnetic interactions and phonon exchange between paramagnetic ions. Our detailed analysis of the nuclear magnetization time-evolution after saturation in the diluted paramagnet has shown that accounting for the stretched exponential time dependence is necessary not only for understanding the short-time recovery, but also to describe long time behavior of the nuclear magnetization. Reasonable agreement is achieved considering the possible sources of errors in the approximations that have been made in the theory of spin diffusion [18,19].

By taking advantage of the simplicity of the system where the main physical parameters can be determined independently with various techniques, we have developed a theoretical description of the $^{19}$F nuclear spin-lattice relaxation which fits the experimental results quite well. The nuclear relaxation rates increase remarkably in the vicinity of the anti-crossings, and our study proves that this is mainly due to direct energy exchange between nuclei and paramagnetic ions. Measurements of the relaxation rates in the region of anti-crossings provide information about random crystal fields and homogeneous broadening of energy levels of impurity ions.

To describe with greater precision the cross-relaxation in the subsystem of paramagnetic ions and its effect on the relaxation rate of host nuclei, it would be necessary to derive a more comprehensive theory of the cross-relaxation, and also to perform measurements of nuclear spin dynamics in the highly diluted samples. However, the current model has allowed us to determine quite adequately the relative importance of the various relaxation mechanisms on nuclear relaxation rate.

This work was supported by INTAS (project 03-51-4943), the Ministry of Education and Science of Russian Federation (project RNP 2.1.1.7348), National Science Foundation grant DMR-0710525, and the QUEMOLNA and MAGMANET European-NMP programs. BZM is grateful to V.A. Atsarkin for very helpful discussions.

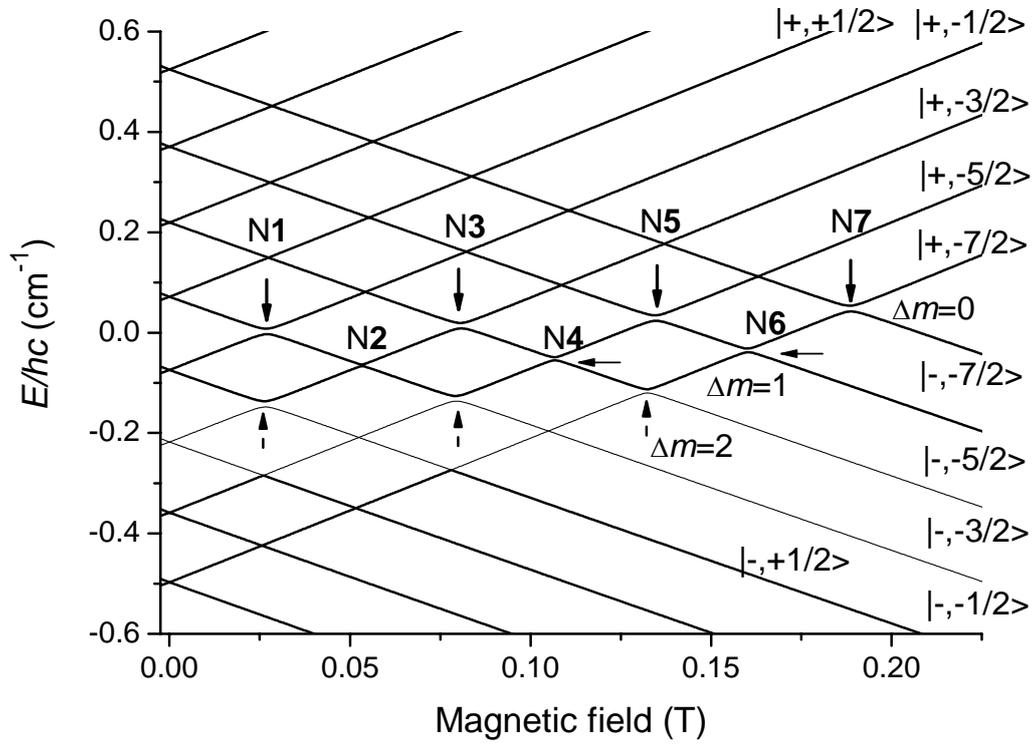

Fig. 1. Electron-nuclear sublevels of the Ho$^{3+}$ ground doublet vs. the external magnetic field strength $B$ ($\theta = 27.5^0$; $\varphi = 45^0$). The wave functions $|\pm, m\rangle$ are specified by the nuclear spin projection $m$.

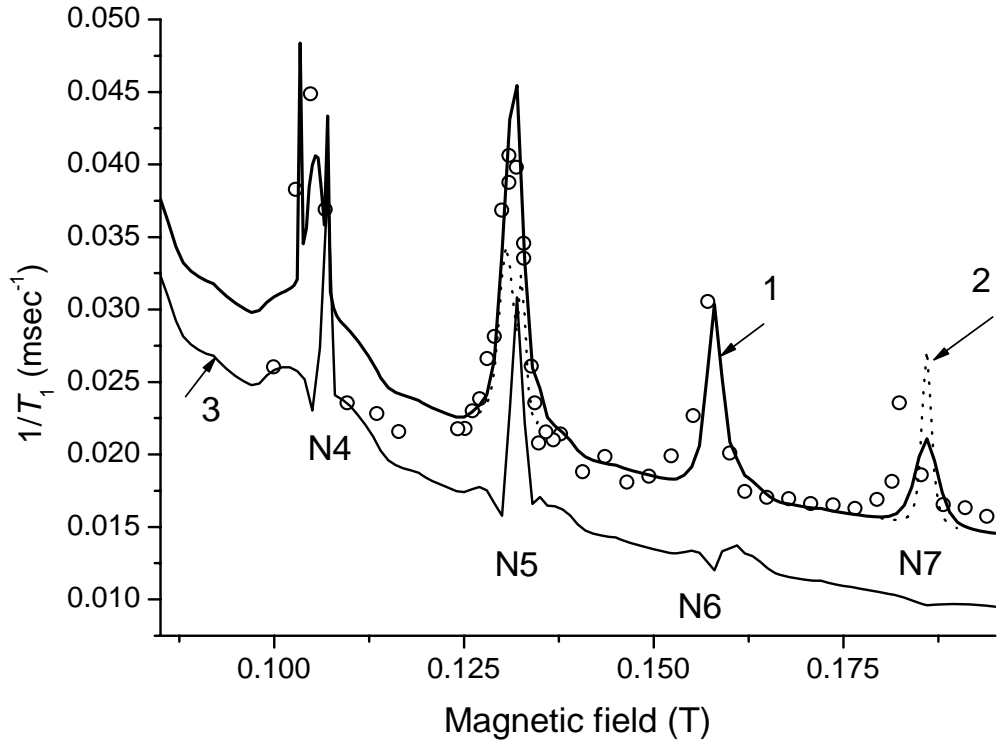

Fig. 2. Nuclear relaxation rates measured at a temperature 1.7 K vs. the external magnetic field, with the field declined from the *c*-axis by $27.5^0$ (symbols). The curves (1) and (2) represent results of simulations with and without accounting for the random crystal field, respectively. The curve (3), shifted down for clarity, is obtained without accounting for the second term in right-hand side of Eq. (7) that describes the inelastic relaxation rates.

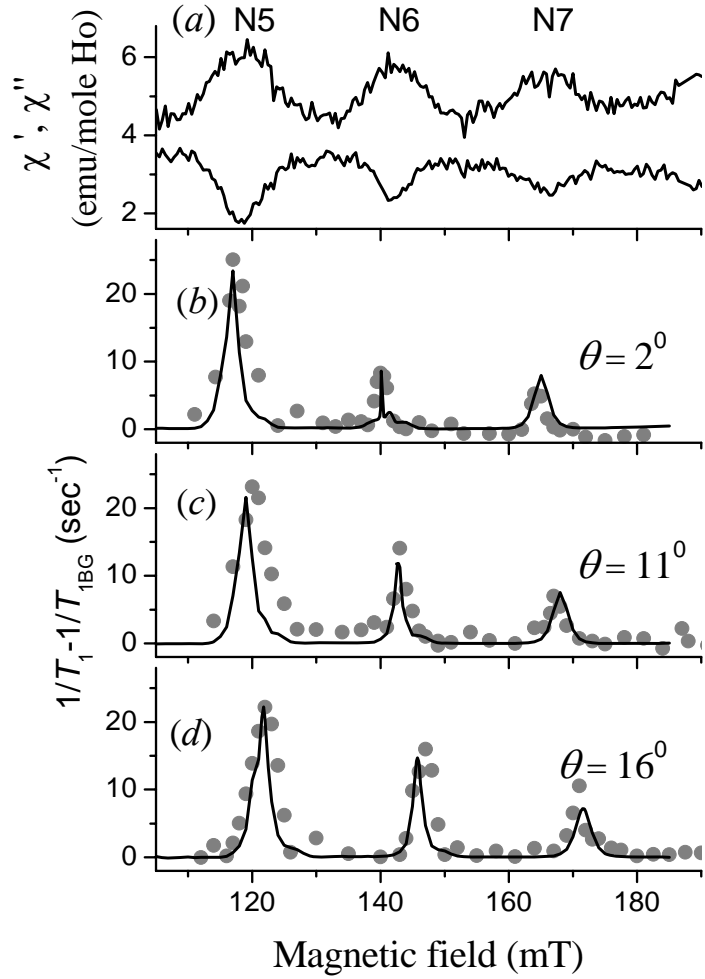

Fig. 3. Measured field dependences of the in-phase (upper curve in panel (a)) and out-of-phase (lower curve in panel (a)) ac-susceptibilities in the LiYF$_4$:Ho$^{3+}$ (0.1 %) sample for collinear dc and ac fields parallel to the *c*-axis at the frequency 350 Hz and temperature 2 K and the fluorine nuclear relaxation rates (with the background term subtracted, panels (b,c,d)) at $T = 1.7$ K for different angles $\theta$ between the magnetic field and the *c*-axis (symbols); solid curves represent results of simulations (see text).

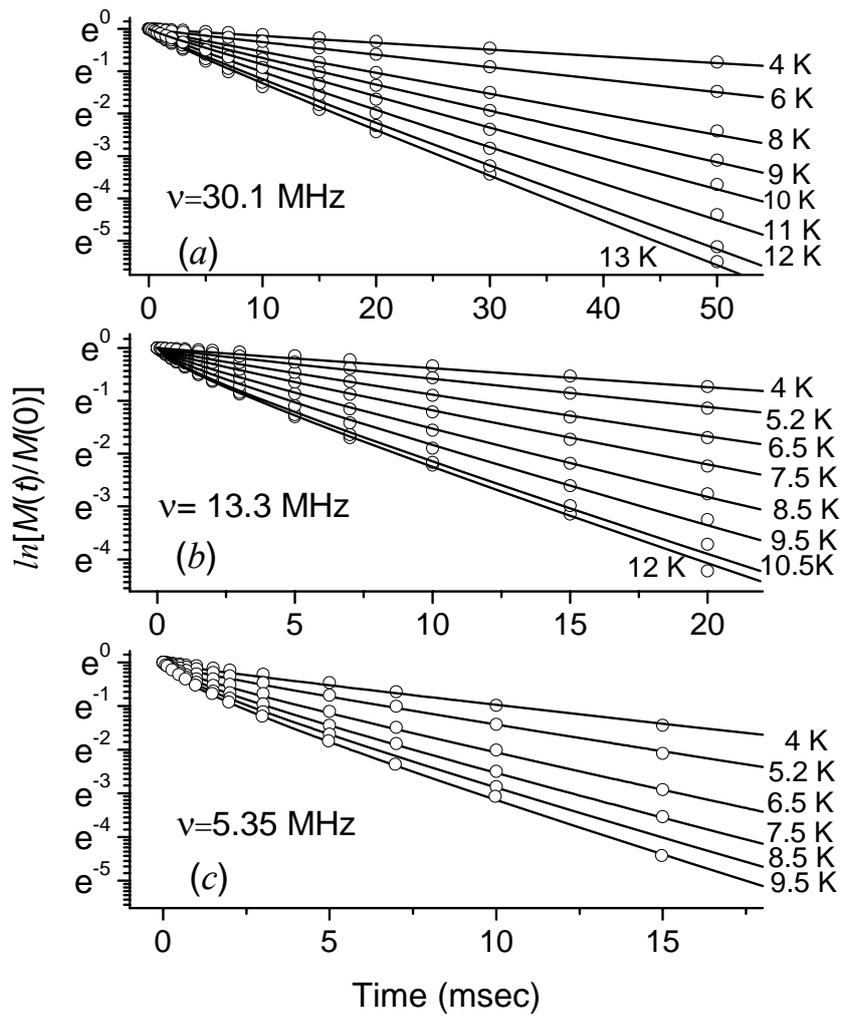

Fig. 4. Time decays of the fluorine nuclear magnetization taken in the magnetic fields $B = 0.751$ T (*a*), 0.332 T (*b*) and 0.1335 T (*c*), declined from the *c*-axis by $27.5^0$ at different temperatures (symbols), $\nu$ is the corresponding NMR frequency. The solid curves represent results of fitting by the function $\exp(-t/T_1-(t/T_s)^{1/2})$, as described in the text.

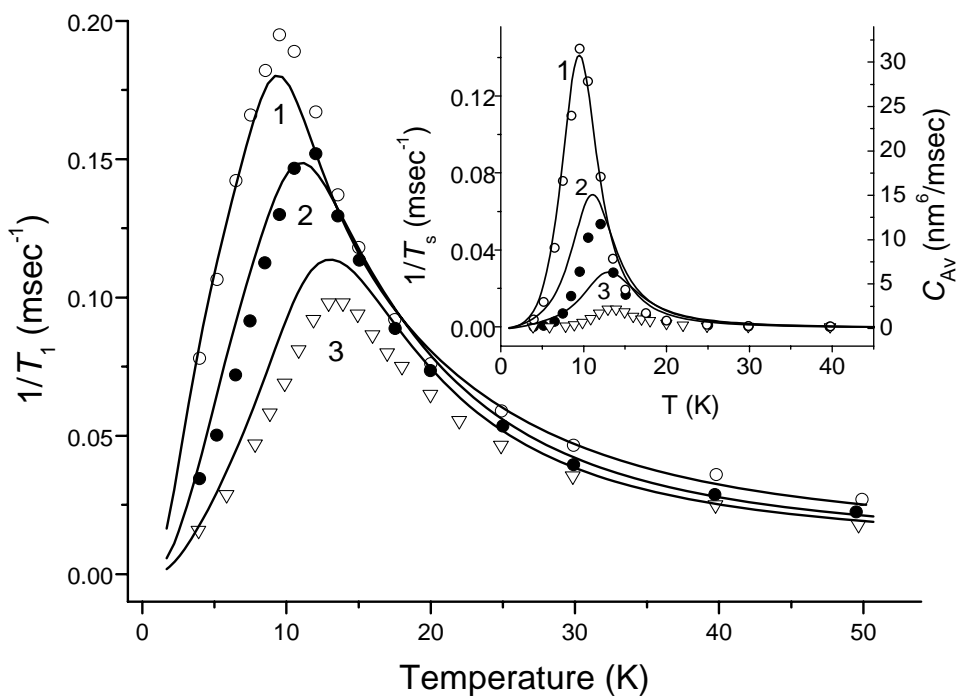

Fig. 5. Measured (symbols) and simulated (curves) temperature dependences of $^{19}$F nuclear relaxation rates in magnetic fields $B$ = 0.1335 T (1), 0.332 T (2) and 0.751 T (3) declined by 27.5$^0$ from the $c$-axis. The solid curves in the insert represent the computed values of $C_{Av}$ vs. temperature (right y-axis).